\documentclass[sigconf,review=false]{acmart}

\usepackage{booktabs} % To thicken table lines

\usepackage[english]{babel}
\usepackage{moresize}
\usepackage{amsmath}
\usepackage{algorithmic}
\usepackage{balance}
\usepackage{comment}
\usepackage{paralist}
\usepackage{bm}
\usepackage{pgfplots}
\usetikzlibrary{pgfplots.dateplot}

\usepackage{flushend}
\usepackage[english]{babel}
\usepackage[latin1]{inputenc}
\usepackage{mathrsfs}
\usepackage{graphicx}

\usepackage{amssymb}
\usepackage{amsfonts}
\usepackage{url}
\usepackage{longtable}
\usepackage{rotating}
\usepackage{multirow}
\usepackage{mathrsfs}
\usepackage{subfigure}
\usepackage{enumitem}
\usepackage[linesnumbered,algoruled,boxed,lined]{algorithm2e}
\usepackage{adjustbox}
\usepackage{hyperref}

\usepackage{pgfplots}
\usetikzlibrary{pgfplots.dateplot}

\usepackage{filecontents}
\definecolor{tblue}{RGB}{31,119,180}
\definecolor{torange}{RGB}{255,127,14}
\definecolor{tgreen}{RGB}{44,160,44}
\definecolor{tred}{RGB}{214,39,40}
\definecolor{tpurple}{RGB}{148,103,189}

\newcommand{\hide}[1]{} %hide

\newcommand{\ie}{\textit{i}.\textit{e}.}
\newcommand{\eg}{\textit{e}.\textit{g}.} 
\newcommand{\wrt}{\textit{w}.\textit{r}.\textit{t}}

\setcopyright{acmcopyright}
\def\model{HGCL}

\keywords{Recommendation, Self-Supervised Learning, Contrastive Learning, Graph Neural Network, Heterogeneous Graph Representation}

\copyrightyear{2023}
\acmYear{2023}
\setcopyright{acmcopyright}\acmConference[WSDM '23]{Proceedings of the Sixteenth ACM International Conference on Web Search and Data Mining}{February 27-March 3, 2023}{Singapore, Singapore}
\acmBooktitle{Proceedings of the Sixteenth ACM International Conference on Web Search and Data Mining (WSDM'23), February 27-March 3, 2023, Singapore, Singapore}
\acmPrice{15.00}
\acmDOI{10.1145/3539597.3570484}
\acmISBN{978-1-4503-9407-9/23/02}


\begin{document}

\fancyhead{}

% \title{Global Context Enhanced Social Recommendation via Graph Mutual Information Maximization}
% \title{Graph Meta Network for Multi-Behavior Recommendation with Interaction Heterogeneity and Diversity}
\title{Heterogeneous Graph Contrastive Learning \\ for Recommendation}

\author{Mengru Chen}
\affiliation{
  \institution{South China University of Technology}
  \city{Guangzhou}
  \country{China}
}
\email{cmr777qyx@gmail.com}

\author{Chao Huang}
\authornote{Chao Huang is the corresponding author.}
\affiliation{
  \institution{University of Hong Kong}
  \city{Hong Kong}
  \country{China}
}
\email{chaohuang75@gmail.com}

\author{Lianghao Xia}
% \authornotemark[1]
\affiliation{
  \institution{University of Hong Kong}
  \city{Hong Kong}
  \country{China}
}
\email{aka_xia@foxmail.com}

\author{Wei Wei}
% \authornotemark[1]
\affiliation{
  \institution{University of Hong Kong}
  \city{Hong Kong}
  \country{China}
}
\email{weiweics@connect.hku.hk}

\author{Yong Xu}
% \authornotemark[1]
\affiliation{
  \institution{South China University of Technology}
  \city{Guangzhou}
  \country{China}
}
\email{yxu@scut.edu.cn}

\author{Ronghua Luo}
% \authornotemark[1]
\affiliation{
  \institution{South China University of Technology}
  \city{Guangzhou}
  \country{China}
}
\email{rhluo@scut.edu.cn}

\begin{abstract}
Graph Neural Networks (GNNs) have become powerful tools in modeling graph-structured data in recommender systems. However, real-life recommendation scenarios usually involve heterogeneous relationships (\eg, social-aware user influence, knowledge-aware item dependency) which contains fruitful information to enhance the user preference learning. In this paper, we study the problem of heterogeneous graph-enhanced relational learning for recommendation. Recently, contrastive self-supervised learning has become successful in recommendation. In light of this, we propose a \underline{H}eterogeneous \underline{G}raph \underline{C}ontrastive \underline{L}earning (\model), which is able to incorporate heterogeneous relational semantics into the user-item interaction modeling with contrastive learning-enhanced knowledge transfer across different views. However, the influence of heterogeneous side information on interactions may vary by users and items. To move this idea forward, we enhance our heterogeneous graph contrastive learning with meta networks to allow the personalized knowledge transformer with adaptive contrastive augmentation. The experimental results on three real-world datasets demonstrate the superiority of \model\ over state-of-the-art recommendation methods. Through ablation study, key components in \model\ method are validated to benefit the recommendation performance improvement. The source code of the model implementation is available at the link https://github.com/HKUDS/HGCL.
\end{abstract}

\maketitle

\section{Introduction}
\label{sec:intro}

In recent years, Graph Neural Networks (GNNs) have become successful in encoding relationships between users and items in recommender systems~\cite{wu2022graph}. The key ideal of GNNs is to learn node (user or item) representations through the aggregation of neighboring feature information across graph propagation layers. However, many GNN-based collaborative filtering (CF) models merely focus on homogeneous interaction relationships in the generated user-item connection graphs~\cite{wang2019neural,chen2020revisiting,wu2021self}. In real-world recommenders, heterogeneous relational information is ubiquitous, such as social network connections between users and knowledge-aware item dependencies with semantic relatedness. In this paper, we address the challenge of incorporating heterogeneous side information into the collaborative filtering for enhancing recommender system.

Inspired by the success of GNNs in a variety of recommendation tasks, researchers attempt to design heterogeneous graph neural networks to embed rich semantics of heterogeneous relations into latent representations. However, the representation power of most existing studies are often hindered by the limitation of sparse training labels. In other words, current heterogeneous graph neural networks are label data-hungry learning models, and thus may not generate quality user/item embeddings with sparse interaction labels for model optimization of recommenders~\cite{wu2021self,long2021social}. 

Contrastive self-supervised learning, emerging as promising representation techniques for addressing data sparsity issue with data augmentation from unlabeled data itself. By integrating contrastive learning with graph neural networks, Graph Contrastive Learning (GCL) has emerged as effective solution to enhance the robustness of learned representations in the absence of sufficient observed labels~\cite{you2020graph} over graph structures. The general idea of GCL is to research the alignment between embeddings encoded from two graph contrastive representation views. In GCL-based self-supervision, the agreement between representations of positive contrastive samples will be maximized, while the distance between embeddings of negative pairs will be pushed away. Motivated by this, we bring the benefits of GCL into the heterogeneous relational learning to improve recommendation performance. 

However, it is non-trivial to effectively realize the heterogeneous relational learning, because the dependencies between side information and user-item interaction modeling are often not monomorphic but diverse in nature. For example, social influence among users may be different due to their personalized characteristics and diverse user-specific interaction pattern. Blindly augmenting the preference learning of users without considering their individual characteristics easily lead to suboptimal representations. In this paper, we investigate the problem of heterogeneous graph learning for recommendation by learning a contrastive augmentor. In essence, we need to solve the challenges in our designed recommender system: i) how to effectively transfer the side knowledge across different views; ii) how to perform heterogeneous relational contrastive learning with personalized augmentation.

To tackle the aforementioned challenges, we propose the principled framework, termed as \underline{H}eterogeneous \underline{G}raph \underline{C}ontrastive \underline{L}earning (\model). Specifically, we first leverage the heterogeneous graph neural network as encoder, the rich semantics of heterogeneous relationships are preserved in the encoded embeddings. To cope with the personalized augmentation, we propose a tailored contrastive learning framework which designs a meta network to encode personalized characteristics of users and items. It allows us to perform user- and item-specific augmentation for transferring informative signals across different relational views. \\\vspace{-0.12in}

\indent The contributions of our work can be summarized as follows:
\begin{itemize}[leftmargin=*]

\item \model\ advances the recommender system with heterogeneous graph contrastive learning, providing a general and universal framework to incorporate heterogeneous side information into recommender under a graph contrastive learning paradigm. \\\vspace{-0.12in}

\item \model\ solves our problem by integrating meta network with contrastive learning for adaptive augmentation to enable user-specific and item-specific knowledge transfer. It advances graph contrastive learning with customized cross-view augmentation. \\\vspace{-0.12in}

\item We conduct extensive experiments on real-world recommendation datasets to validate that our \model\ framework is capable of significantly improving performance over other strong baselines.

\end{itemize}

%\clearpage
% \input{model}
\section{Related Work}
\label{sec:relate}

\subsection{GNN-based Recommender Systems}
In general, Graph Neural Networks (GNNs) follow the idea of message passing across different graph layers by consisting of information propagation and aggregation. Under graph neural architecture, many GNN-based recommender systems are proposed to capture various graph-structured relationships in recommendation. For example, GNNs are adopted for modeling user-item interaction graph for generating latent representations via cross-layer information propagation in NGCF~\cite{wang2019neural}, LR-GCCF~\cite{chen2020revisiting} and SHT~\cite{xia2022self}. To enhance collaborative relation learning with social influence among users, social relation encoders in some existing studies are also built upon graph neural networks, such as GraphRec~\cite{fan2019graph}, KCGN~\cite{huang2021knowledge} and MHCN~\cite{yu2021self}. Furthermore, graph neural networks have become effective solution to encode sequential patterns of item sequences for time-aware recommendation, including SURGE~\cite{chang2021sequential} and MA-GNN~\cite{ma2020memory}. In recent years, modeling multiple graph connections with GNNs (\eg, MBGCN~\cite{jin2020multi} and MGNN~\cite{zhang2020multiplex}) has attracted much attention in handling more complex recommendation scenario with diverse user behaviors. In those GNN-based multi-behavior recommenders, the behavior-aware message passing is considered to reflect diverse user preference from multi-behavior data. There also exist some multimedia recommender systems (\eg, GRCN~\cite{wei2020graph}, DualGNN~\cite{wang2021dualgnn}) built upon graph neural networks to incorporate multi-modal information into recommendation.

\subsection{Contrastive Learning for Recommendation}
Recently, the contrastive self-supervised learning has been noticed by researchers. It is because the generated self-supervision signals can be used to enrich user representation learning. In recommender systems, contrastive learning can be a powerful tool to incorporate self-supervision signals for data augmentation with the alignment between contrastive representation views. For example, many studies aim to address the data sparsity issue in recommenders by proposing various graph augmentation schemes for embedding contrasting, \eg, SGL~\cite{wu2021self}, HCCF~\cite{xia2022hypergraph} and NCL~\cite{lin2022improving}. In particular, random node/edge dropout operations are adopted for generate graph contrastive learning views in SGL~\cite{wu2021self}. In HCCF~\cite{xia2022hypergraph}, local-global contrastive learning is designed for self-supervised augmentation based on parameterized hypergraph structures. In those contrastive graph CF models, the embedding uniformity can be improved based on InfoNCE-based contrasting. There also exist some studies leveraging contrastive learning in knowledge graph representation in recommender systems, such as KGCL~\cite{yang2022knowledge} and KGIC~\cite{zou2022improving}. In addition, contrastive learning has been used in various recommendation scenarios, including sequential recommendation~\cite{wang2022multi}, multi-behavior recommendation~\cite{wei2022contrastive}, and multi-interest recommendation~\cite{zhang2022re4}. In this work, a novel heterogeneous graph contrastive learning paradigm is proposed to fill the gap in recommender system by capturing heterogeneous relationships in recommendation with contrastive learning.

\subsection{Heterogeneous Graph Learning}
Heterogeneous graphs is ubiquitous in real-life applications with various types of nodes and connections. Representation learning over heterogeneous graphs aims to encode node embeddings in which the rich semantics with relation heterogeneity can be well preserved~\cite{yang2020heterogeneous}. To achieve this goal, heterogeneous graph neural networks become the promising techniques to provide state-of-the-art representation results. For example, HAN~\cite{wang2019heterogeneous} enhances the graph attention network with the capability of dealing with heterogeneous types of nodes and relations based on meta-path construction. Motivated by the transformer framework, HGT~\cite{hu2020heterogeneous} designs a graph transformer network to enable the heterogeneous message passing using self-attention to calculate the propagation weights between nodes. In addition, both intra- and inter-metapath aggregation are considered in MAGNN~\cite{fu2020magnn} to fuse information from different meta-paths over heterogeneous graphs. In HGIB~\cite{yang2021heterogeneous}, information bottleneck is extended to heterogeneous graph learning with self-supervision among homogeneous graphs. Towards this research line, this paper tackles an important but unexplored task of heterogenous graph contrastive learning recommendation.
\section{Methodology}
\label{sec:solution}
\begin{figure*}

    \centering
    \includegraphics[width=\textwidth]{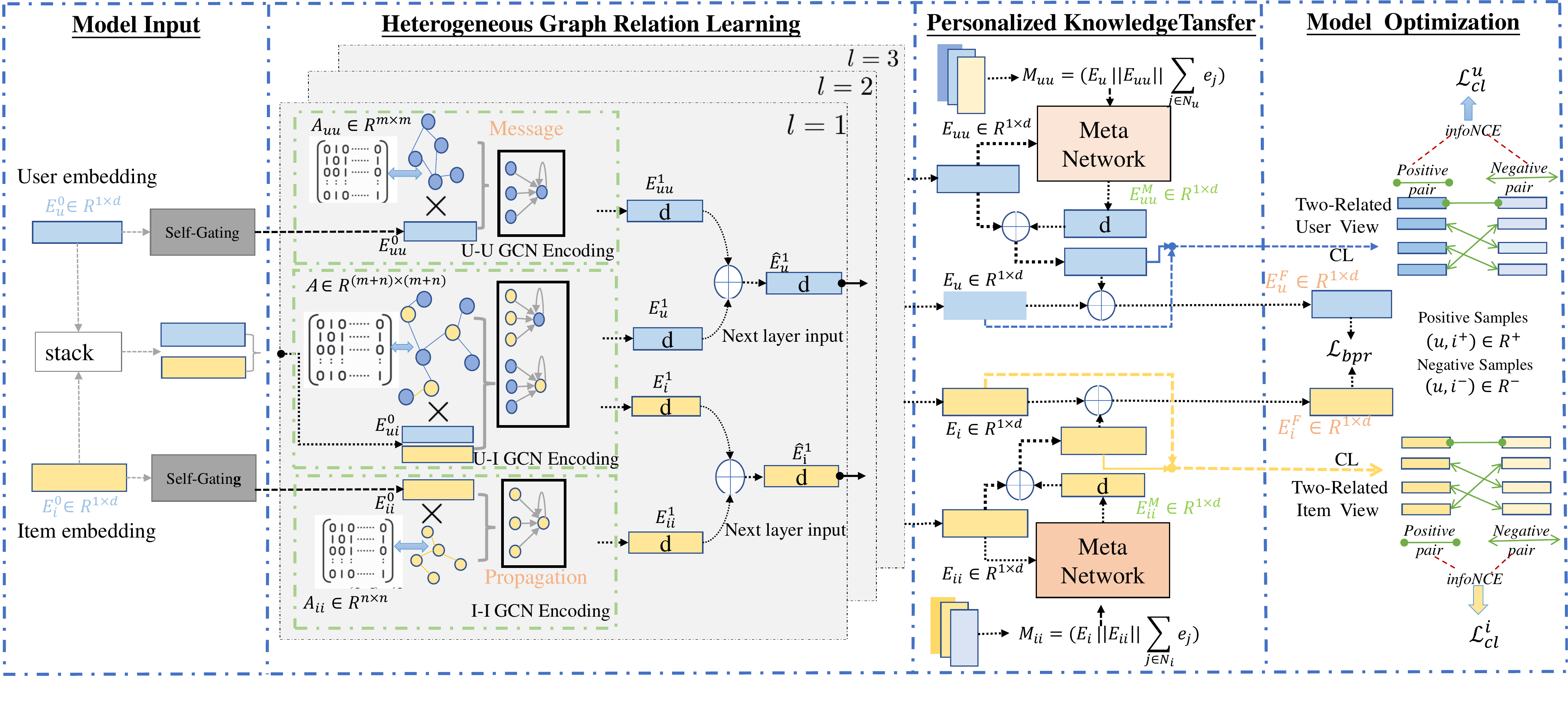}
     \vspace{-0.1in}
     \caption{The model flow of the proposed \model\ framework. \model\ includes three key components:  (1) Heterogeneous graph representation extraction and fusion by heterogeneous graph neural network on user-user graph, user-item graph and item-item graph. (2) Meta network for personalized cross-view dependencies modeling between the auxiliary views and the interaction view. (3) Jointly parameter optimization with adaptive contrastive learning between the heterogeneous relational views. }
    \label{fig:framework}
\end{figure*}

In this section, we elaborate the model design of our proposed \model\ framework, which enhances representation learning on heterogeneous graph for recommendation with cross-view contrastive learning. The overall framework of \model\ is illustrated in Figure~\ref{fig:framework}.

\subsection{Preliminaries}
Relations in real-life recommender systems are often heterogeneous to contain diverse semantic information from users and items. We represent the user-item interaction data with the graph $\mathcal{G}_{ui}=\{\mathcal{V}_{u}, \mathcal{V}_{i}, \mathcal{E}_{ui}\}$, where $\mathcal{V}_{u}$ and $\mathcal{V}_{i}$ denote the sets of users and items, respectively. In graph $\mathcal{G}_{ui}$, if user $u$ has adopted item $i$, then there exists an edge between $u$ and $i$ ($(u,i)\in\mathcal{E}_{ui}$). To represent social relationships among users, graph $\mathcal{G}_{uu}=\{\mathcal{V}_u,\mathcal{E}_{uu}\}$ is defined to include user-wise social connections with the edge set $\mathcal{E}_{uu}$. To incorporate item-wise relations, we define the item graph $\mathcal{G}_{ii}=\{\mathcal{V}_i, \mathcal{E}_{ii}\}$ to connect dependent items with external knowledge (\eg, item category). For these defined graphs, we define three adjacent matrices $\textbf{A}_{ui}\in\mathbb{R}^{m\times n}, \textbf{A}_{uu}\in\mathbb{R}^{m\times m}$ and $\textbf{A}_{ii}\in\mathbb{R}^{n\times n}$, corresponding to graph $\mathcal{G}_{ui}$, $\mathcal{G}_{uu}$ and $\mathcal{G}_{ii}$, respectively. Here, $m$ and $n$ denotes the number of users and items, respectively. The objective of this work is to predict unobserved interactions between users and items given the graphs with relation heterogeneity.

% We represent the user-item interaction data with a bipartite graph $\mathcal{G}_{ui}=\{\mathcal{V}_{u}, \mathcal{V}_{i}, \mathcal{E}_{ui}\}$, where $\mathcal{V}_{u}$ and $\mathcal{V}_{i}$ denote the sets of users and items, respectively. If a user $u$ has adopted an item $i$, then there exists an edge $(u,i)\in\mathcal{E}_{ui}$. Analogously, we use the user-wise graph $\mathcal{G}_{uu}=\{\mathcal{V}_u,\mathcal{E}_{uu}\}$ to encode the social relations between users. If two items $i_1$ and $i_2$ belong to the same category, then an item-item relation $(i_1, i_2)$ is included by the item-wise graph $\mathcal{G}_{ii}=\{\mathcal{V}_i, \mathcal{E}_{ii}\}$. For convenience, we use adjacent matrices $\textbf{A}_{ui}\in\mathbb{R}^{m\times n}, \textbf{A}_{uu}\in\mathbb{R}^{m\times m}$ and $\textbf{A}_{ii}\in\mathbb{R}^{n\times n}$ to encode the information of the above three graphs $\mathcal{G}_{ui}, \mathcal{G}_{uu}$ and $\mathcal{G}_{ii}$, respectively, where $m$ and $n$ correspond to the number of users and items, respectively. The objective of heterogeneous graph-based recommendation is to make predictions $\hat{y}_{u,i}$ on the un-observed user-item relations $(u,i)$, based on the observed heterogeneous graph $\{\mathcal{G}_{ui}, \mathcal{G}_{uu}, \mathcal{G}_{ii}$\}.

\subsection{Heterogeneous Graph Relation Learning}
\subsubsection{\bf Relation-Aware Embedding Initialization}
To encode the heterogeneous collaborative relations with the modeling of high-order connectivity, we employ heterogeneous graph neural networks to learn embeddings from the user-item graph $\mathcal{G}_{ui}$, user-user graph $\mathcal{G}_{uu}$, and item-item graph $\mathcal{G}_{ii}$. To begin with, we assign id-corresponding embeddings $\textbf{e}_u, \textbf{e}_i\in\mathbb{R}^d$ initialized by xavier initializer~\cite{glorot2010understanding}, where $d$ denotes the hidden dimensionality. The node-specific embeddings form the initial embedding matrices $\textbf{E}_u^0\in\mathbb{R}^{m\times d}$ and $\textbf{E}_i^0\in\mathbb{R}^{n\times d}$. The initial embeddings are fed into different graph encoders for user-item domain, user-user domain, and item-item domain. To highlight the differences in interactive patterns between the three relation types, we train a self-gating module~\cite{yu2021self} to derive the relation-aware embeddings for user-wise social connections and item-wise semantic relations from the common initial embedding space, which are showned as follows:
\begin{align}
\begin{split}
    \textbf{E}_{uu}^0 = \textbf{E}_u^0\odot \sigma(\textbf{E}_u^0\textbf{W}_{g}+\textbf{b}_{g}) ;\quad 
    \textbf{E}_{ii}^0 =
    \textbf{E}_i^0\odot\sigma(\textbf{E}_i^0\textbf{W}_{g}+\textbf{b}_{g})
    \end{split}
\end{align}
where $\textbf{E}_{uu}^0\in\mathbb{R}^{m\times d}$ and $\textbf{E}_{ii}^0\in\mathbb{R}^{n\times d}$ are the embeddings for the homogeneous graphs $\mathcal{G}_{uu}$ and $\mathcal{G}_{ii}$ for user-wise and item-wise relations, respectively. $\sigma(\cdot)$ denotes the sigmoid activation function. $\odot$ denotes element-wise multiplication operation. $\textbf{W}_g \in  \mathbb{R}^{d \times d}$ and $\textbf{b}_g \in \mathbb{R}^{d \times 1}$ are the transformation and bias parameters. Through the self-gating mechanism with multiplicative skip-connection~\cite{dauphin2017language}, embeddings $\textbf{E}_{uu}^0, \textbf{E}_{ii}^0$ not only share common semantic with initial embeddings $\textbf{E}_u^0,\textbf{E}_i^0$ for user-item interactions, but also gain the flexibility to characterize the user-user and item-item relations.

\subsubsection{\bf Heterogeneous Message Propagation}
Among the above initial embedding matrices, $\textbf{E}_u^0, \textbf{E}_i^0$ are used as input for the user-item view, $\textbf{E}_{uu}^0$ and $\textbf{E}_{ii}^0$ are used as input for the user-user view and the item-item view, respectively. We first apply a graph convolutional neural network as the encoder for three views of graph structures. Without loss of generality, we elaborate the modeling for user-item relation graph as an example. Specifically, given the user-item interaction graph $\mathcal{G}_{ui}$, our \model\ iteratively refine the user and item embeddings with the message propagation as follows:
% The three views are modeled with a common graph convolutional neural architecture. Without loss of generality, we elaborate the modeling for user-item domain as an example. Given the user-item interactive graph $\mathcal{G}_{ui}$, our \model\ iteratively refine the user embeddings and item embeddings to encode high-order information as follows:
\begin{align}
    \textbf{e}_u^{l+1} = \sum_{i \in \mathcal{N}_u}
    \frac{1}{\sqrt{|\mathcal{N}_u|}\sqrt{|\mathcal{N}_i|}}
    \textbf{e}_i^l;\quad
    \textbf{e}_i^{l+1} = \sum_{u \in \mathcal{N}_i}
    \frac{1}{\sqrt{|\mathcal{N}_i|}\sqrt{|\mathcal{N}_u|}}
    \textbf{e}_u^l
\end{align}
\noindent where $\mathcal{N}_u$ and $\mathcal{N}_i$ denote the neighbor set of target nodes $u$ and $i$, respectively. $\textbf{e}_u^l, \textbf{e}_i^l\in\mathbb{R}^d$ denotes the embedding vectors of user $u$ and item $i$ in the $l$-th iteration. $\textbf{e}_u^0, \textbf{e}_i^0$ are row vectors of the embedding matrices $\textbf{E}_u^0, \textbf{E}_i^0$, respectively. Inspired by the effectiveness and efficiency of lightweight GCN~\cite{he2020lightgcn} in CF recommendation, our relation-aware message passing paradigm is configured without transformation and non-linear activation. Analogously, the embeddings $\textbf{E}_{uu}^l$ for user-user graph and the embeddings $\textbf{E}_{ii}^l$ for item-item graph are refined iteratively following the same GCN schema.

\subsubsection{\bf Heterogeneous Information Aggregation}
Inspired by the \emph{soft meta-path} design in~\cite{hu2020heterogeneous}, the information in each iteration is aggregated from heterogeneous relations. Through multiple iterations of heterogeneous message propagation, the high-order embeddings preserve heterogeneous semantics with multi-hop connections. In particular, the embddings of users and items are updated through the following defined heterogeneous fusion procedure:
\begin{align}
    \widehat{\textbf{E}}_{u}^{l+1} = f(\textbf{E}_{u}^{l+1}, \textbf{E}_{uu}^{l+1}); \quad
    \widehat{\textbf{E}}_{i}^{l+1} = f(\textbf{E}_{i}^{l+1}, \textbf{E}_{ii}^{l+1}) 
\end{align}
\noindent 
\noindent where the refined embeddings in the $l+1$ iteration $\widehat{\textbf{E}}_u^{l+1} \in \mathbb{R}^{m \times d}$, $\widehat{\textbf{E}}_i^{l+1} \in \mathbb{R}^{n \times d}$ integrate heterogeneous semantics and become the input for the next layer. $f$ denotes the heterogeneous information fusion function. Here, to reduce the model complexity, we use element-wise mean pooling as the fusion function $f(\cdot)$.

To further aggregate heterogeneous information with encoded layer-specific representations ($1\leq l \leq L$), we generate the overall embeddings of users and items as follows:
\begin{align}
    \textbf{E}_{u} = \textbf{E}_{u}^0 + \sum_{l=1}^L{ \frac{\textbf{E}_{u}^{l}}{||\textbf{E}_{u}^{l}||} }; \quad
    \textbf{E}_{i} = \textbf{E}_{i}^0 + \sum_{l=1}^L{ \frac{\textbf{E}_{i}^{l}}{||\textbf{E}_{i}^{l}||} }
    % \textbf{E}_{u} = \textbf{E}_{u}^0 + \sum_{l=1}^L{ \frac{\textbf{E}_{u}^{b}}{||\textbf{E}_{u}^{l}||} }\\
    %         \textbf{E}_{u} = \textbf{E}_{u}^0 + \sum_{l=1}^L{ \frac{\textbf{E}_{u}^{b}}{||\textbf{E}_{u}^{l}||} }
\end{align}
\noindent where $L$ denotes the maximum number of GCN iterations. The output of each GCN layer are normalized. We add the initial embeddings $\textbf{E}_u^0, \textbf{E}_i^0$ using skip connections. The above presented formulas indicate the layer-specific representation aggregation for the user-item interaction view. The embeddings of user-user social view (\ie~$\textbf{E}_{uu}$) and the item-item dependency view (\ie~$\textbf{E}_{ii}$) are obtained through multi-order information aggregation in an analogous way.

% the heterogeneous embeddings for the user-user social view (\ie~$\textbf{E}_{uu}$) and the embeddings for the item-item view (\ie~$\textbf{E}_{ii}$) are obtained in an analogous way.

\subsection{Cross-View Meta Network}
Our \model\ aims to enhance the collaborative filtering by incorporating the heterogeneous relational knowledge from both user social connections and item external dependence. However, in real-life user modeling scenario, the influence of user and item side information over the user-item interaction patterns may be different among users. For example, some users are more likely to be influenced by the recommendations from their social friends, while others often adopt items based on their own preference. Therefore, it is necessary to perform personalized knowledge transfer from side information to guide the learning of user-specific preference. Towards this end, we design a cross-view meta network to enable the customized knowledge distillation from both user and item side.

% Our \model\ aims to improve collaborative filtering for the user-item relation prediction by incorporating the auxiliary user-user social relations and the item-item similarity relations. The interactive patterns of the three views are individually captured in the aforementioned heterogeneous relation learning module. In this section, our \model\ is motivated to fill the potential gap between the three views, so as to better utilize the auxiliary views for user and item embeddings refinement. Specially, the cross-view connections between the different domains can be very distinct for different users and items. For example, users who spend more time handling social relations are more likely to pay attention to items favored by their social ties. To empower our \model\ with the capability of personalized cross-view transferring, we employ a cross-view meta network schema to adaptively distill accurate representations from the auxiliary views to the target collaborative domain.

\subsubsection{\bf Meta Knowledge Extraction}
To generate personalized mapping from the auxiliary views (user and item side information) to the encoding of user-item interaction for each user and each item, we first extract meta knowledge to preserve important features of users and items \wrt~both the auxiliary views and interaction view. Specifically, the distilled meta knowledge for the user-user relation view and the item-item relation view is obtained as follows:
\begin{align}
    \textbf{M}_{uu}=\textbf{E}_u||\textbf{E}_{uu}||\sum_{i \in \mathcal{N}_u} \textbf{e}_i;\quad
    \textbf{M}_{ii}=\textbf{E}_i||\textbf{E}_{ii}||\sum_{u \in \mathcal{N}_i} \textbf{e}_u
\end{align}
\noindent where $\textbf{M}_{uu}\in\mathbb{R}^{m\times 3d}, \textbf{M}_{ii}\in\mathbb{R}^{n\times 3d}$ represent the meta knowledge that encodes the context information to generate personalized knowledge transfer functions for user and item side knowledge, respectively. Motivated by~\cite{xia2021graph}, the meta knowledge contains the node representation $\textbf{E}_{uu}, \textbf{E}_{ii}$ of the source domains (\ie~the user-user and item-item relation view) as well as the embeddings of the target user-item interaction view $\textbf{E}_u, \textbf{E}_i$. In addition, we incorporate the neighborhood information into the meta knowledge. Specifically, the embeddings of the auxiliary domains characterize the users' social influence and item semantic relatedness. The embeddings of the user-item view captures the item-related interactive patterns of users. The additional neighborhood information explicitly enhances the modeling of direct graph connections. By collectively considering the three dimensions of information, the meta knowledge is able to well-reflect the important contextual signals for personalized cross-view knowledge transferring.

\subsubsection{\bf Personalized Cross-View Knowledge Transfer}
In our \model, the extracted meta knowledge is utilized to generate a parameterized knowledge transfer network with customized transformation matrices. The proposed meta neural network is
% Then the extracted meta knowledge is handled by the meta network to generate weight matrices customized for each user and each item. The customized weights adaptively transform the embeddings from the auxiliary-view representation space to the target user-item-view representation space. The proposed meta neural network is formulated as:
\begin{equation}
    \begin{split}
    % f_{L}(\textbf{M}_{uu})=
    \left\{\ 
    \begin{aligned} f^1_{mlp}(\textbf{M}_{uu})  \rightarrow  \textbf{W}_{uu}^{M1}\\ f^2_{mlp}(\textbf{M}_{uu})  \rightarrow  \textbf{W}_{uu}^{M2}\\
    \end{aligned}
    \right.
    \end{split}
\end{equation}
\noindent where $f^1_{mlp},f^1_{mlp}$ are meta knowledge learner consisting of two fully-connected layers with PReLU activation function. The functions take the meta knowledge $\textbf{M}_{uu}$ as input, and output the customized transformation matrices $\textbf{W}_{uu}^{M1}\in\mathbb{R}^{m\times d\times k}, \textbf{W}_{uu}^{M2}\in\mathbb{R}^{m\times k\times d}$. Both two parameter tensors contain $m$ matrices for each of the $m$ users. The customized transformations are generated according to the unique characteristics of the corresponding users and items to realize the personalized knowledge transfer. The two sets of matrices restrict the rank of the transformation to $k<d$, which not only reduces the number of trainable parameters of meta knowledge learnder and enhance the model stability. Inspired by the personalized bridge function in~\cite{zhu2021personalized}, we leverage the generated parameter matrices and a non-linear mapping function to build our customized transfer network as follows:
\begin{equation}
    \begin{split}
    % f_{L}(\textbf{M}_{uu})=
     \textbf{E}_{uu}^M = \sigma( \textbf{W}_{uu}^{M1}\textbf{W}_{uu}^{M2} \textbf{E}_{uu}) 
    \end{split}
\end{equation}
\noindent where $\sigma(\cdot)$ denotes the PReLU activate function. $\textbf{E}_{uu}^M \in \mathbb{R}^{m\times d}$ contains the embeddings transformed by the customized mapping function for the user-user social view. Then the customized embeddings are utilized to enhance the user embeddings encoded from the user-item interactions. The fusion process for users is conducted by the following weighted summation:
\begin{equation}
    \begin{split}
    \textbf{E}_u^{F} = \alpha_u*\textbf{E}_u + (1-\alpha_u)*(\textbf{E}_{uu} + \textbf{E}_{uu}^M);\ 
    % \textbf{E}_i^{F} = \alpha_i*\textbf{E}_i + (1-\alpha_i)*\textbf{E}_{ii}^M
    \end{split}
\end{equation}
\noindent where $\alpha_u\in\mathbb{R}$ denotes the hyperparameter which controls the weight between the user-item interaction view embedding and the user-user social view embedding. Here the original embeddings of user-user relation view is also utilized for better optimization. $\textbf{E}_u^F\in\mathbb{R}^{m\times d}$ represent the final embeddings used for the main task of recommendation. The foregoing process elaborates the calculation for cross-view user embedding customization. The cross-view item embeddings $\textbf{E}_{ii}^M, \textbf{E}_{i}^F$ can be generated in a similar way.

% and the item embeddings $\textbf{E}_{ii}^M, \textbf{E}_{i}^F$ are obtained in the same way. 

\subsection{Heterogeneous Relational Contrastive Learning for Augmentation}
\subsubsection{\bf Cross-View Contrastive Learning}
To further enhance the representation learning of our \model\ framework with more supervision signals to mitigate the data sparsity issue, we design the cross-view contrastive learning paradigm to enhance the robustness of the heterogeneous relational learning with self-augmentation. Concretely, the embeddings of the two auxiliary views (\ie~$\textbf{E}_{uu}^M$ and $\textbf{E}_{ii}^M$) are aligned with the embeddings of the user-item interaction view (\ie~$\textbf{E}_u$ and $\textbf{E}_i$). With this design, the embeddings of the auxiliary views serve as effective regularization to influence the user-item interaction modeling with the self-supervised signals.

% are regularized to be distributed more similarly as the target view. And the embeddings of the target view is also refined by users' social ties and items' category information.

To capture the diverse user preference by considering the personalized cross-view knowledge transfer, we integrate the personalized cross-view knowledge transfer with the contrastive learning in our recommender system. In particular, the cross-view embedding alignment is conducted in an adaptive way between different representation views. The auxiliary-view-specific embeddings $\textbf{E}_{uu}, \textbf{E}_{ii}$ are processed by the personalized mapping functions generated by the meta network, to yield the personalized auxiliary embeddings $\textbf{E}_{uu}^M, \textbf{E}_{ii}^M$. The meta network is trained to filter noisy features in the auxiliary views to match the user-item interaction view.

% In particular, the information of the auxiliary views may be biased \wrt~the target view. For example, two items may be close to each other in our item-item view, as they share the same category. But such items may attract very different users because of their differences in price and brand. To mitigate the inherent inconsistency between the auxiliary view and the target view, the aforementioned cross-view meta network is also employed to adaptively align the auxiliary views and the target view in our contrastive learning module. To be specific, the auxiliary-view-specific embeddings $\textbf{E}_{uu}, \textbf{E}_{ii}$ are processed by the personalized mapping functions generated by the meta network, to yield the personalized auxiliary embeddings $\textbf{E}_{uu}^M, \textbf{E}_{ii}^M$. The meta network is trained to filter noisy features in the auxiliary views to match the target user-item interactive view.

\subsubsection{\bf InfoNCE-based Contrastive Loss}
With the help of our heterogeneous graph relation learning and cross-view meta networks, we obtain two sets of embeddings for both users and items, \ie~$\textbf{E}_{uu}^M, \textbf{E}_u$ for users, and $\textbf{E}_{ii}^M, \textbf{E}_i$ for items. The embeddings are obtained via encoding the user-item interaction data, and the user/item-side auxiliary knowledge. Inspired by the success of recent contrastive self-supervised learning in recommendation~\cite{wu2021self, xia2022hypergraph}, we propose to empower the user/item representation learning of our \model\ method with the InfoNCE-based contrastive learning loss between two representation views as follows:
\begin{equation}
    \begin{split}
    \mathcal{L}^u_{cl} = \sum_{u \in \mathcal{V}_u}{-\log \frac{\exp\left( s( \textbf{e}^M_{uu}+\textbf{e}_{uu}, \textbf{e}_u )/ \tau\right)}{\sum_{u' \in \mathcal{V}_u}{\exp\left( s( \textbf{e}^M_{uu}+\textbf{e}_{uu}, \textbf{e}_u')/ \tau\right)}}}
    \end{split}
\end{equation}
\noindent where $\textbf{e}^M_{uu} \in \mathbb{R}^{d}$, $\textbf{e}_u \in \mathbb{R}^{d}$ are the embedding vectors from the matrices $\textbf{E}_{uu}^M$ and $\textbf{E}_u$, respectively. $s(\cdot)$ denotes the similarity function, which can be inner product or cosine similarity. Here we use cosine similarity as our $s(\cdot)$. $\tau$ represents the temperature coefficient, which is capable of automatically identifying difficult negative samples. $u'$ indicates negative samples with different indices. Analogously, we can obtain the InfoNCE loss $\mathcal{L}^i_{cl}$ of items aspect. Finally, the total contrastive loss is
$\mathcal{L}_{cl}$ = $\alpha_1$ $\ast$$\mathcal{L}^u_{cl}$ + $\alpha_2$ $\ast$$\mathcal{L}^i_{cl}$, where $\alpha_1$ and $\alpha_2$ denote two hyperparameters for weight tuning.

\subsection{Optimization Objectives of \model}
With the fused embeddings $\textbf{E}_u^F, \textbf{E}_i^F$, our \model\ forecast the likelihood of user $u$ interacting with item $i$ via dot-product: $\hat{y}_{u,i}=\textbf{e}_u^{F\top}\textbf{e}_i^{F}$, where $\textbf{e}_u^F$ and $\textbf{e}_i^F$ denote the final embedding vectors of user $u$ and item $i$ from the fused embedding matrices. $\hat{y}_{u,i}\in\mathbb{R}$ denotes the score that indicates the likelihood of user $u$ interacting with item $i$. Larger $\hat{y}_{u,i}$ reflects larger probability of interaction. To optimize our \model\ with the recommendation task, we follow recent works and adopt the Bayesian Personalized Ranking (BPR)~\cite{rendle2012bpr} pair-wise loss function. Specifically, each training sample is configured with a user $u$, a positive item $i^+$ that the user has interacted with, and a negative item $i^-$ that the user has not interacted with. For each training sample, we maximize the prediction score as follows:
\begin{align}
    \mathcal{L}_{bpr} = \sum_{(u,i^+,i^-)\in O} - \ln (\text{sigmoid} (\hat{y}_{u,i^+} - \hat{y}_{u,i^-})) + \lambda || \Theta ||^2 
\end{align}
\noindent where $\ln(\cdot)$ and $\text{sigmoid}(\cdot)$ denote the logarithm function and the sigmoid function, respectively. $\lambda$ denotes a hyperparameter to determine the weight of the regularization term. Combining the BPR loss function with the augmented cross-view contrastive learning loss, the overall training loss is presented as follows:
\begin{equation}
    \begin{split}
    \mathcal{L} = \mathcal{L}_{bpr} + \beta*\mathcal{L}_{cl}
    \end{split}
\end{equation}

% \subsubsection{\textbf{Model Complexity Analysis}}
% \noindent \textbf{Model Complexity Analysis}.
\subsection{Model Complexity Analysis}
We give detailed analysis on the time complexity of our \model\ model to measure the efficiency of our method. The heterogeneous GNN module of \model\ employs a lightweight network structure, which takes $\mathcal{O}((|\mathcal{E}_{ui}|+|\mathcal{E}_{uu}|+|\mathcal{E}_{ii}|)\times d\times $L$)$ time. In the cross-view meta network, the highest computational cost comes from the meta network for personalized mapping function generation, which takes $\mathcal{O}((m+n)\times d^2\times k)$ time. In the heterogeneous relational contrastive learning component, $\mathcal{O}(b\times (m+n)\times d)$ time complexity is needed in each batch (batch size $d$) to calculate the InfoNCE loss across the heterogeneous relational views. Overall, the above discussed first and third components are identical to the complexity of state-of-the-art self-supervised GNN recommendation methods (\eg, SGL~\cite{wu2021self}). The second module takes the complexity which is close to the complexity of a vanilla GNN as $k$ is typically small.

% In conclusion, our \model\ achieves model efficiency in terms of theoretical complexity analysis.
\section{Evaluation}
\label{sec:eval}

\begin{table*}[t!]
\caption{Performance comparison of all methods on different datasets in terms of \emph{NDCG} and \emph{HR}.}
\vspace{-0.1in}
\centering
% \scriptsize
%\footnotesize
%\small
\setlength{\tabcolsep}{0.5mm}
% \resizebox{\textwidth}{0.5mm}
\small
\begin{tabular}{|c|c|c|c|c|c|c|c|c|c|c|c|c|c|c|c|c|c|}
\hline
Data                   & Metric & SAMN   & DGRec  & ETANN  & NGCF & KGAT   & MKR    & GraphRec & DANSER & HERec  & MCRec  & HAN  &HeCo& HGT& MHCN  & SMIN         & \textbf{\model} \\ \hline
\multirow{2}{*}{Ciao}     & H@10      &  0.6576 & 0.6653 & 0.6738 & 0.6945 & 0.6601 & 0.6793 & 0.6825   & 0.6730 & 0.6800 & 0.6772 & 0.6589 &
0.6867 &
0.6939&
0.7053
&0.7108& \textbf{0.7376} \\ \cline{2-18} 
                          & N@10    &  0.4561 & 0.4953 & 0.4665 & 0.4894 & 0.4512 & 0.4589 & 0.4730   & 0.4521 & 0.4712 & 0.4708&0.4469&0.4867&0.4869&0.4928 & 0.5012 &  \textbf{0.5261} \\ \hline
\multirow{2}{*}{\begin{tabular}[c]{@{}c@{}}Epinions\end{tabular}}  & H@10      & 0.7592 & 0.7603 & 0.7650 & 0.7984 & 0.7510 & 0.7647 & 0.7723   & 0.7714 & 0.7642 & 0.7630 & 0.7505 &0.7998&0.8150&
0.8201 &
0.8179&
\textbf{0.8367} \\ \cline{2-18} 
& N@10    &  0.5614 & 0.5668 & 0.5663 & 0.5945 & 0.5578 & 0.5669 & 0.5751   & 0.5741 & 0.5495 & 0.5326 & 0.5275&0.5910&0.6126&0.6158 &0.6137& \textbf{0.6413} \\ \hline
\multirow{2}{*} {Yelp}& H@10      &  0.7910 & 0.7950 & 0.8031 & 0.8265 & 0.7881 & 0.8005 & 0.8098   & 0.8077 & 0.7928 & 0.7869 & 0.7731&0.8359&0.8364&0.8344& 0.8478 & \textbf{0.8712} \\ \cline{2-18} 
& N@10    &  0.5516 & 0.5593 & 0.5560 & 0.5854 & 0.5501 & 0.5635 & 0.5679   & 0.5692 & 0.5612 & 0.5590 & 0.5604&0.5847&0.5883&0.5799 &0.5993&  \textbf{0.6310} \\ \hline
\end{tabular}
\label{tab:overall_performance}
%\vspace{-0.1in}
\end{table*}

\begin{table}[t]
    \caption{Statistics of experimented datasets}
\vspace{-0.1in}
    \label{tab:data}
    \centering
    % \footnotesize
    %\scriptsize
    \small
	\setlength{\tabcolsep}{1mm}
\begin{tabular}{lllll}
\hline
Dataset  & User \# & Item \# & Interaction \# & Sparsity  \\ \hline
Ciao     & 6776    & 101415  & 265308         & 99.9614\% \\
Epinions & 15210   & 233929  & 630391         & 99.9823\% \\
Yelp     & 161305  & 114852  & 957923         & 99.9948\% \\ \hline
\end{tabular}
% \vspace{-0.1in}
\end{table}

In this section, we perform model evaluation to investigate the effectiveness of our \model\ and baseline methods. We also analyze the impact of key modules and model robustness. Our experiments are designed to address the following research questions:
\begin{itemize}[leftmargin=*]

\item \textbf{RQ1}: How does \model\ perform compared with existing methods? \\\vspace{-0.1in}

\item \textbf{RQ2}: Is it beneficial to incorporate key components in our \model\ to boost the recommendation performance? \\\vspace{-0.1in}

\item \textbf{RQ3}: How doe \model\ perform in different environments with varying sparsity degrees of user interaction data? \\\vspace{-0.1in}

\item \textbf{RQ4}: How does key hyperparameters affect model performance?

\end{itemize}

\subsection{Experimental Settings}

\subsubsection{\bf Datasets} In our experiments, our \model\ framework is evaluated on
three real-world datasets from online platforms. We present the data statistics in Table~\ref{tab:data} and present the details of each dataset as followed. \textbf{Ciao and Epinions}. They are two benchmark recommendation datasets collected from online review systems to contain user rating behaviors over different items. The heterogeneous relations are generated from the contained user and item side information, such as user trust relationships and item categorical information. \textbf{Yelp}. This dataset contains heterogeneous relations (\eg, user social relations, venue rating behaviors, business attributes) in the recommendation scenario of local businesses on Yelp platform.

\subsubsection{\bf Baselines} To evaluate the validity of our proposed method, we compare \model\ with various systems for comprehensive performance comparison. The baseline details are described as below.
\begin{itemize}[leftmargin=*]

\item \textbf{SAMN}~\cite{chen2019social}: This model designs attention-based memory network to consider the difference of social influence among users for improving the user-item interaction modeling. \\\vspace{-0.1in}

\item \textbf{DGRec}~\cite{song2019session}: This approach utilizes recurrent neural network to model dynamic interests of users and graph attention network to model social influence for recommendation. \\\vspace{-0.1in}

\item \textbf{ETANN}~\cite{chen2019efficient}: It designs an adaptive transfer scheme from the social domain to the encoding process of user-item interaction patterns by considering user-user relationships. \\\vspace{-0.1in}

\item \textbf{NGCF}~\cite{wang2019neural}: We incorporate the social information among users into the representative GNN-based collaborative filtering model. The message passing is built based on graph convolutions. \\\vspace{-0.1in}

\item \textbf{KGAT}~\cite{wang2019kgat}: In this baseline, the item knowledge-based relationships are incorporated into the graph attention mechanism for enhancing recommender system. \\\vspace{-0.1in}

\item \textbf{MKR}~\cite{wang2019multi}: It utilizes knowledge graph as the side information to assist the recommendation with multi-task learning framework. Different tasks are associated with cross and compress units. \\\vspace{-0.1in}

\item \textbf{GraphRec}~\cite{fan2019graph}: This method jointly models the user-user social graph and user-item interaction graph to reflect the relation heterogeneity in recommendation. \\\vspace{-0.1in}

\item \textbf{DANSER}~\cite{wu2019dual}: This recommender system learns two-fold of social effects with user-specific and dynamic attentive weights estimated via contextual multi-armed bandit. \\\vspace{-0.1in}

% Various social effects are fused with different weights generated via contextual multi-armed bandit. \\\vspace{-0.1in}

\item \textbf{HERec}~\cite{shi2018heterogeneous}: It aims to encode heterogeneous information in recommendation based on meta-path-based random walk. \\\vspace{-0.1in}

\item \textbf{MCRec}~\cite{hu2018leveraging}: Co-attention mechanism is proposed to capture the heterogeneous relationships in recommender system. \\\vspace{-0.1in}

\item \textbf{HAN}~\cite{wang2019heterogeneous}: We apply this representative heterogeneous graph neural network to generate user and item representations via meta-path-based attention encoder. \\\vspace{-0.1in}

\item \textbf{HGT}~\cite{hu2020heterogeneous}: It introduces heterogeneous mutual attention for message passing scheme to refine user/item embeddings along with diverse relations in the heterogeneous graph structures. \\\vspace{-0.1in}

\item \textbf{HeCo}~\cite{wang2021self}: It is a self-supervised method which integrates contrastive learning with heterogeneous GNNs to consider local and high-order graph structures. Embeddings encoded with different meta-path-based connections are used for contrasting. \\\vspace{-0.1in}

\item \textbf{SMIN}~\cite{long2021social}: It is a self-supervised social recommender system which incorporates auxiliary graph learning task into the main task to improve the recommendation performance. \\\vspace{-0.1in}

\item \textbf{MHCN}~\cite{yu2021self}: In this recommender, a multi-channel hypergraph convolutional network is designed to consider global relationships among users based on motifs.

% \item \textbf{SimGCL}~\cite{yu2022graph}: It is a graph contrastive learning method which performs embedding-level augmentation for recommendation.

\end{itemize}

\begin{figure*}[t!]
    \centering
    \includegraphics[width=1.8\columnwidth]{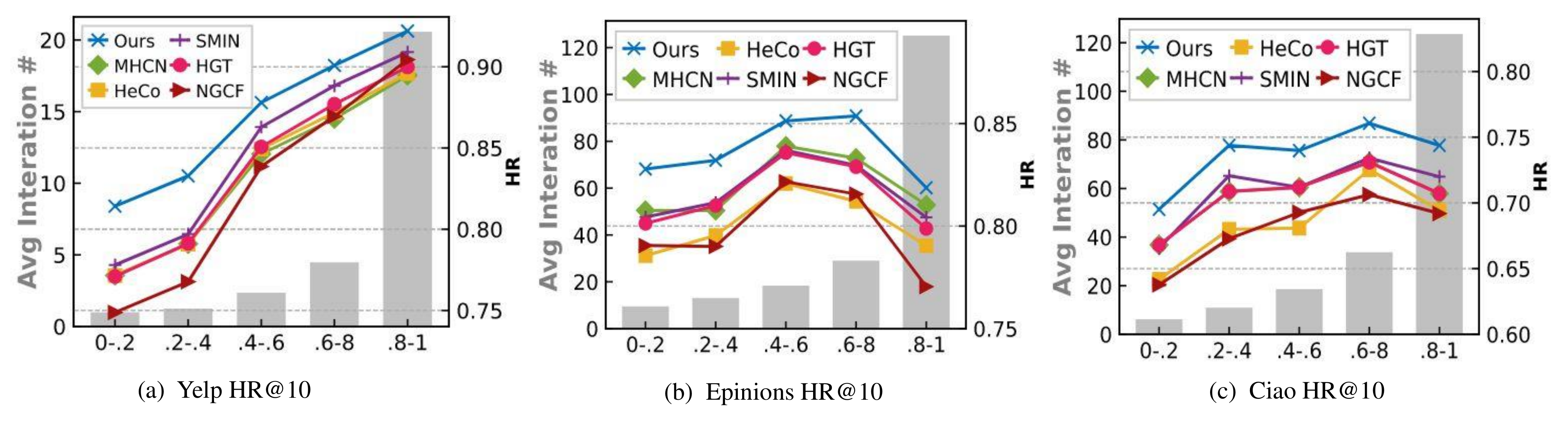}
    \vspace{-0.1in}
    %   \captionsetup{font={small,bf}, justification=raggedright} 
    \caption{Performance comparison with respect to different data sparsity degrees on three datasets.}
    % \caption{citeseer multi level ablation}
    % \vspace{-0.1in}
    \label{fig:sparsity}
\end{figure*}

\subsubsection{\bf Hyperparameter Settings}
Our \model\ model is implemented using PyTorch. The model is optimized with Adam for parameter learning. In the model implementation, the batch size and learning rate is searched from \{1024, 2048, 4096, 8192\} and \{4e-2, 4.5e-2, 5e-2, 5.5e-2, 6e-2\}, respectively. The embedding size is tuned from the range of \{8, 16, 32, 64, 128\}. The number of graph neural network layers is selected from \{1, 2, 3\}. Additionally, the coefficient $\beta$ of contrastive loss is selected from \{0.2, 0.25, 0.3, 0.35, 0.55, 0.6, 0.65\}. The dimension of low rank matrix decomposition of meta knowledge extraction is chosen from \{1, 2, 3, 4, 5\}.

In our evaluation settings, one positive (interacted) item and 99 negative (non-interacted) items are sampled for each user for performance evaluation. To measure the recommendation accuracy of different methods, two widely-adopted metrics HR(Hit Ratio) and NDCG (Normalized
Discounted Cumulative Gain) are used.

\begin{table}
% \label{tab:1}
\caption{Ablation study on key components of HGCL}
\vspace{-0.1in}
\label{tab:ablation}
\centering
\small
% \includegraphics[width=0.5\textwidth , right]
% \resizebox{\linewidth}{!}{
\begin{tabular}{c|cc|cc|cc}
\hline
Data                                                & \multicolumn{2}{c|}{Ciao}         & \multicolumn{2}{c|}{Epinions}     & \multicolumn{2}{c}{Yelp}          \\ \hline
Metric                                              & HR              & NDCG            & HR              & NDCG            & HR              & NDCG            \\ \hline
w/o-cl                                              & 0.7124          & 0.5015          & 0.8176          & 0.6166          & 0.8471          & 0.6030          \\ \hline
w/o-meta                                            & 0.7215          & 0.5135          & 0.8247          & 0.6282          & 0.8585          & 0.6218          \\ \hline
% M-mlp & 0.7117          & 0.5006          & 0.8112          & 0.6091          & 0.8480          & 0.6070          \\ \hline
w/o-ii                                              & 0.7116          & 0.5055          & 0.8245          & 0.6317          & 0.8573          & 0.6188          \\ \hline
w/o-uu                                              & 0.7149          & 0.5047          & 0.8285          & 0.6266          & 0.8533          & 0.6208          \\ \hline
HGCL                                               & \textbf{0.7376} & \textbf{0.5261} & \textbf{0.8367} & \textbf{0.6413} & \textbf{0.8712} & \textbf{0.6310} \\ \hline
\end{tabular}
\vspace{-0.1in}
\end{table}

\subsection{Performance Comparison (RQ1)}
Table~\ref{tab:overall_performance} reports the performance of all compared methods on different datasets for item recommendation. From the evaluation results, we summarize the following key observations:
\begin{itemize}[leftmargin=*]

\item Our \model\ consistently achieves the significant performance improvement compared with state-of-the-arts. We attribute these improvements to the design of heterogeneous graph contrastive learning: (1) \model\ allows recommender system to perform effective knowledge transfer among heterogeneous relationships to help model user preference; (2) the adaptive contrastive learning has great ability to improve recommendation performance with self-supervision signals between heterogeneous relation views.\\\vspace{-0.12in}

\item Heterogeneous graph neural network-based methods (\eg, HeCo, HGT, SMIN) often offer better performance than other alternative approaches (\eg, SAMN, KGAT, DANSER), which justifies the effectiveness of incorporating heterogeneous relational knowledge of social influence and item semantic relatedness from user and item side into the recommender system. \\\vspace{-0.12in}

\item As can be seen, the observed superior performance of MHCN and SMIN indicates the rationality of augmenting user-item interaction encoding with self-supervised learning technique. The performance gap between our \model\ and those self-supervised learning-enhanced recommenders validates that adaptive self-supervised signal distillation indeed boosts the performance with contrastive personalized knowledge transfer.

\end{itemize}

\subsection{Ablation Study (RQ2)}
We conduct ablation study to validate that the consideration of customized contrastive learning heterogeneous relationships is essential and benefit the performance, as elaborated below:
\begin{itemize} [leftmargin=*]

\item \emph{\textbf{w/o-meta}}: We do not include the meta network in \model\ to allow the personalized knowledge transfer in our developed contrastive learning augmentation across heterogeneous relational views.\\\vspace{-0.12in}

\item \emph{\textbf{w/o-cl}}: We disable the contrastive learning in our model to capture the cross-view dependency between the auxiliary information and user-item interaction modeling.\\\vspace{-0.12in}

\item \emph{\textbf{w/o-ii}}: In this variant, we do not include the item-item graph $\mathcal{G}_{ii}$ to capture the knowledge-aware dependency among items for guiding the learning process of user preference.\\\vspace{-0.12in}

\item \emph{\textbf{w/o-uu}}: In this variant, we do not include the user-user graph $\mathcal{G}_{uu}$ to consider the social influence among users to help encode the user-item interaction patterns.

\end{itemize}

The recommendation performance of \model\ framework and compared variants are presented in Table~\ref{tab:ablation}. In all cases, the performance of \model\ is superior to w/o-cl, reflects the rationalities of our heterogeneous graph contrastive learning for effective augmentation with cross-view knowledge transfer. w/o-meta performs worse than \model\ on different datasets. This result is consistent with our assumption that user/item-specific customized knowledge transfer is helpful to learn user representations. \model\ achieves consistent gain over w/o-ii and w/o-uu, which implies the necessity of considering heterogeneous side information into the recommender system to guide the encoding of user preference.

\begin{figure}
    \centering
 \includegraphics[width=1\columnwidth]{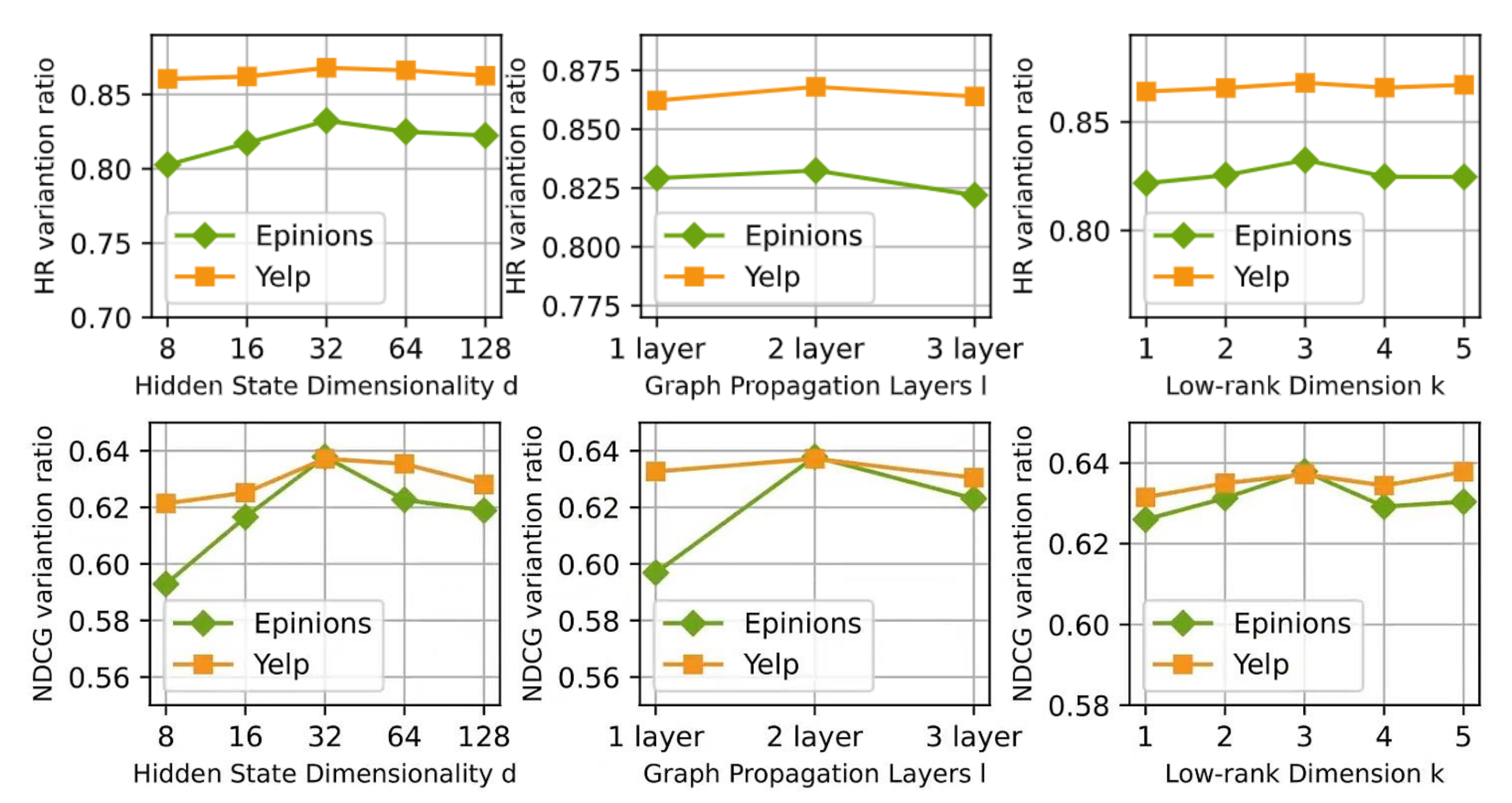}
    \vspace{-0.15in}
    \caption{Hyperparameter study of the \model.}
    % \caption{citeseer multi level ablation}
    \vspace{-0.1in}
    \label{fig:parameter}
\end{figure}

\begin{figure*}[t]
    \centering
    \includegraphics[width=2.1\columnwidth]{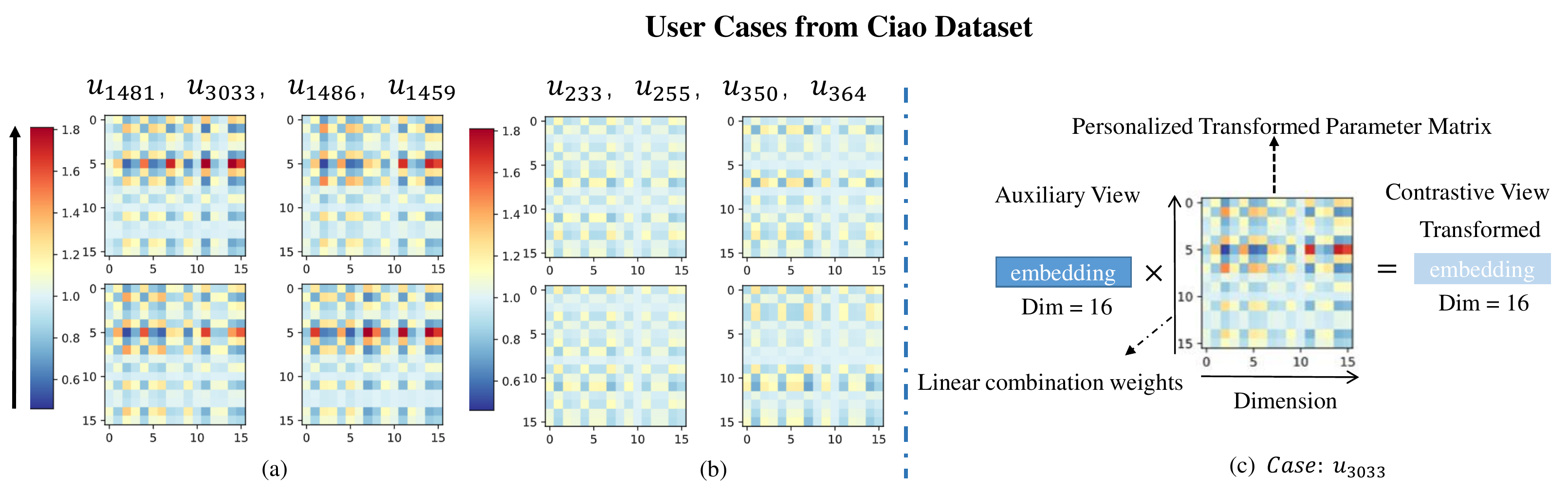}
    \vspace{-0.2in}
    % \captionsetup{font={small,bf}, justification=raggedright}
    % \caption{Case study on the explainability of \model\ . The visualization of   Personalized transformed weights. That parameter matrix  help for embedding transformation from auxiliary view to target view. }
    \caption{Case study on Ciao dataset to visualize the learned contrastive transformation matrices sampled from different users to reflect the diverse social influence. (a): Four users who are more likely to be influenced by their social relations; (b): Four users who are less likely to be influenced by their social relations. (c): The embeddings generated from auxiliary view will be transformed for representation contrasting for self-supervision augmentation of user-item interaction modeling.}
    % \caption{citeseer multi level ablation}
    % \vspace{-0.1in}
    \label{fig:case_study}
\end{figure*}

\subsection{Performance varying Data Sparsity}
In this section, we evaluate the performance of different methods when varying the data sparsity degrees of user interaction data. We divide the set of users into five groups to represent diverse user active degrees. The performance comparison results between \model\ and several baselines are shown in Figure~\ref{fig:sparsity}. The recommendation accuracy of each method is presented in the right side of y-axis with lines. The left side y-axis represents the number of average number of interactions in each user group with bars. It is obvious to see the superior performance of our method under different sparsity environments. The improvements of \model\ may come from the contrastive learning-enhanced cross-view knowledge transfer, because it can effectively capture the user-specific social influence and item-specific semantic relatedness. Therefore, through the conducted experiments, \model\ is able to maintain a decent performance even with sparse user-item interactions.

\subsection{Hyperparameter Analysis}
We further perform parameter sensitivity analysis to show the impact of hidden state dimensionality, the number of graph propagation layers, and low-rank dimension. The results are shown in Figure~\ref{fig:parameter}. From the results, we make the following conclusions.

\begin{itemize}[leftmargin=*]

\item \textbf{Hidden State Dimensionality}. The hidden state dimensionality $d$ is selected from 8 to 128. We can notice that the model performance firstly increases and then reaches saturation when $d=32$. Hence, properly enlarging the embedding dimension size can boost the recommendation performance, but not always being performance gain due to model overfitting. \\\vspace{-0.12in}

\item \textbf{The Number of Graph Propagation Layers}. In graph neural architecture, the number of propagation layers is searched from 1 to 3. The curves depicts that the model achieves the better performance by stacking two layers. This suggests that more layers could capture the high-order neighbors and semantic information. However, deeper GNN architecture can lead to the model over-smoothing and induce noise to the feature representation. \\\vspace{-0.12in}

\item \textbf{Low-rank Decomposition Dimension}. We can observe that the parameter study on the low-rank decomposition dimension $k$ indicates that the best performance is obtained with $k=3$. Smaller value of $k$ may not be sufficient to learn the complex transformation information.

\end{itemize}

\subsection{Qualitative Evaluation}
In our evaluation, we perform case studies on Ciao dataset to visualize the learned personalized contrastive transformation matrix ($\mathbb{R}^{16\times 16}$) to reflect the diverse influence between the auxiliary view (\eg, social relationships) and the user-item interaction view. In Figure~\ref{fig:case_study}, we sample four users who are more (\eg, $u_{1481}$, $u_3033$)/less (\eg, $u_{233}$, $u_{255}$) likely to be influenced by social relationships when adopting items. The corresponding personalized contrastive transformation matrices of different users are visualized to capture diverse knowledge transfer between the social view and interaction view. We can observe that larger values in the learned contrastive transformation matrix indicate larger social influence for this user. With the integration of meta network and contrastive learning, the adaptive contrastive data augmentation can be realized based on the personalized characteristics of users.
% \vspace{-0.1in}
\section{Conclusion}
\label{sec:conclusion}

In this paper, we study the problem of graph representation learning for recommendation with the consideration of heterogeneous relations. To solve this problem, a novel heterogeneous graph contrastive learning model (\model) is proposed to transfer knowledge from side information to the user-item interaction modeling in an adaptive way. In our \model, we propose to identify the informative heterogeneous relations to augment collaborative filtering paradigm. Our experiments on real-world datasets validate that our \model\ outperforms state-of-the-arts by a large margin. In-depth analysis validates the robustness of our model in alleviating data sparsity. One interesting direction for future work is to explore and disentangle the real interest and conformity, by incorporating heterogeneous relationships in recommender systems to alleviate the popularity bias from noisy interaction data of users. Furthermore, in future work, it is also interesting to explore confounding effects for heterogeneous relational learning in recommender systems.

%\clearpage

% \section*{Acknowledgments}
% %We thank the anonymous reviewers for their constructive feedback and comments. 
% This research work is supported by the research grants from the Department of Computer Science \& Musketeers Foundation Institute of Data Science at the University of Hong Kong.

\clearpage

\bibliographystyle{ACM-Reference-Format}
\balance
\bibliography{sigproc.bib}

% \clearpage
% \input{appendix}

\end{document}